\documentclass[twoside]{ilcws07}
\usepackage[latin1]{inputenc}
\usepackage[dvips]{graphicx,epsfig,color}
\usepackage{wrapfig,rotating}
\usepackage{amssymb,amsmath,array}

\def\Journal#1#2#3#4{{#1} {\bf #2}, #3 (#4)}

\def\NPB{{\em Nucl. Phys.} B}
\def\PLB{{\em Phys. Lett.}  B}
\def\PRL{\em Phys. Rev. Lett.}
\def\PRD{{\em Phys. Rev.} D}

\def\beq{\begin{equation}}
\def\eeq{\end{equation}}
\def\bea{\begin{eqnarray}}
\def\eea{\end{eqnarray}}

\begin{document}

\begin{flushright}
Freiburg-PHENO-07/04.
\end{flushright}
\vskip 1.0cm
\begin{center}
{\Large \bf A cosmotopological relation for 
% the gauge group and representations of
 a unified field theory}
\vskip 0.4cm
J.~J.~van~der~Bij\\
\vskip 0.3cm
Insitut f\"ur Physik, Albert-Ludwigs Universit\"at Freiburg \\
H. Herderstr. 3, 79104 Freiburg i.B., Deutschland\\
\end{center}
\vskip 0.7cm

\begin{abstract}
I present an argument, based on the topology of the universe, why there
are three generations of fermions. The argument implies a preferred unified gauge
group of $SU(5)$, but with $SO(10)$ representations of the fermions.
The breaking pattern $SU(5) \rightarrow SU(3)\times SU(2)\times U(1)$
is preferred over the pattern $SU(5) \rightarrow SU(4)\times U(1)$.
On the basis of the argument one expects an asymmetry in the early universe
microwave data, which might have been detected already.
\end{abstract}
\vskip 0.7cm

\section{Introduction}
The standard model based on the gauge group $SU(3)\times SU(2) \times U(1)$
with its complicated set of fermions, constrained by the anomaly is
simply asking for a unification into a larger group. It is well known
that such a unification is possible and quite natural. Given
the fact that there is no convincing evidence for the existence of
extra $Z's$ or $W's$ the known vector bosons point towards a unification
within a group of rank $4$, namely $SU(5)$ \cite{georgi}.\

However, at the latest with the discovery of neutrino masses, it has become clear
that the natural unification for the fermions is within the group 
$SO(10)$ \cite{fritzsch},
since each generation forms an irreducible spinor representation of $SO(10)$.
So naively speaking the vector bosons and the fermions point toward a different
form of unification. Of course the situation can be described through the breaking
of the symmetry with a number of Higgs fields, but one would hope for
a more fundamental explanation for this feature. Another fact of phenomenology is
the existence of precisely three generations of fermions. It is natural to wonder
whether the group question $SU(5)$ versus $SO(10)$  is related to  the 
question of the number of generations.
We are therefore looking for an argument to constrain the representation content
and the gauge group of the theory. The only type of argument known that can give
such constraints is based on some form of an anomaly. As anomalies are intimately 
related to topology, one is led to the question: what is the topology
of space and time?\\

In typical Robertson-Walker metrics the topology of the world is a sphere
or an open space. In higher dimensional cosmologies many shapes are possible.
For instance one can take $M_4 \times U^n(1)$, a torus shape for the higher dimensions.
One would like to describe the universe as starting in a higher dimensional space,
where some of the dimensions dynamically shrink to become too small to 
be seen today. This idea is however difficult to realize in practice, as
the Einstein equations tend to lead to an expansion of the universe in all
directions. Therefore we start with the opposite assumption. We assume that
the early universe was lower dimensional than it is now. To be more 
precise we assume a topology $M_3 \times U(1)$ for space-time, where the
radius of the third dimension is small in the early universe. 
Through the cosmological expansion the radius becomes very large, so that 
at the present epoch one finds a very large, essentially flat and isotropic
universe. This behaviour can be realized in the Bianchi-I universes \cite{kasner}.
For a modern review on non-isotropic and/or non-homogeneous
cosmological models, see \cite{wain}.\\

Since the present universe is very large and flat, we have no 
direct information on its overall topology.
However an indirect indication of the topology can be found
 through the appearance of matter fields, while not all matter fields
can lead to a consistent quantum field theory for a given topology.
For instance, in order to have spin-1/2 fields the Stiefel-Whitney
class of the manifold has to vanish. Another example appears in three 
spacetime dimensions when a Chern-Simons mass term for gauge fields is present.
For non-abelian gauge fields, invariance under large gauge transformations requires
the mass to be quantized in units of the coupling constant[5-10]. When massive fermions are 
present, loop effects give rise to a finite renormalization of the Chern-Simons
term\cite{rao}, leading to restrictions on the number of fermion fields. For massless fermions
the restrictions are unchanged, but then they come from a 
non-perturbative parity anomaly[11-14]. 
Closely related to the parity anomaly and of interest to our argument
is the CPT anomaly\cite{frans1}, that can appear if the universe has a preferred
direction, for instance when it has the topology $M_3\times U(1)$. In this
case a Chern-Simons like term can arise for the photon, due to the presence
of Weyl-fermions in the theory.
\beq
L_{CS\,like} = m_{ph}~n^{\alpha} \epsilon_{\alpha \beta \gamma \delta}A^{\beta}F^{\gamma \delta}
\eeq
Here $n^{\alpha}$ is the preferred direction in space and $m_{ph}$ is the
mass-like term for the photon.
This term violates Lorentz invariance and CPT. It is  called Chern-Simons like,
because a true Chern-Simons term exists only in three dimensions.
If this term is present, it gives rise to a number of interesting effects
in the propagation of photons in spacetime.
The CPT anomaly is actually a lifting to four dimensions
of the three dimensional parity anomaly. The connection is most easily seen
when one takes the radius of the $U(1)$ to zero, thereby dimensionally
reducing the theory. In this limit the four dimensional photon becomes a three
dimensional Chern-Simons photon. The four dimensional Weyl-fermions
become three dimensional Dirac fermions. The structure of the anomaly can therefore
be analysed in the dimensionally reduced theory. For a didactic introduction to the
issues involved we refer to \cite{frans2}.\\

\section{The argument}
Since we assume the early universe to be of the form $M_3 \times U(1)$,
with  a very small compactified $U(1)$ in the early universe, we  are
interested in the consistency of the quantum field theory, reduced to three
dimensions, ignoring the compactified dimension. Quantum anomalies in
such space-times have been discussed before within the subject of 
Chern-Simons theories. In such theories Chern-Simons terms are a part
of gravitational and Yang-Mills fields. However their coefficients are quantized.
The matter fields in the theory give rise to loop-induced gravitational
Chern-Simons terms, that in general will not satisfy the correct quantization
rules. Thereby one can constrain the matter fields of the theory.
In [17-20] corrections to the gravitational Chern-Simons term
due to fermions were considered, in \cite{pisarski, yang}  the effects of the
vector bosons were considered. The back-reaction of a gravitational Chern-Simons term 
on the vector bosons was considered in \cite{lerda}.

The gravitational action in three dimensions contains two terms.
One is the ordinary Einstein Lagrangian:
\beq
L = -(1/\kappa^2)\sqrt{g} R
\eeq
where as usual, R is the curvature scalar, $g_{\mu\nu}$ is the metric tensor,
$g$ the determinant of the metric and $\kappa^2$ is Newton's constant.
To this action a Chern-Simons term can be added:
\beq
L_{CS} = -\frac{i}{4\kappa^2\mu}\epsilon^{\mu\nu\lambda}
(R_{\mu\nu ab} \omega_{\lambda}^{ab} + \frac{2}{3}\omega_{\mu a}^b
\omega_{\nu b}^c \omega_{\lambda c}^a).
\eeq
where
\beq
R_{\mu\nu ab}  = \partial_{\mu} \omega_{\nu ab}
+\omega_{\mu a}^c \omega_{\nu cb} -(\mu \leftrightarrow \nu)
\eeq
is the curvature tensor and $\omega_{\mu ab}$ is the spin connection.
The gravitational Chern-Simons charge
\beq
q_{gr} = \frac {6 \pi}{\mu \kappa^2}
\eeq
is quantized and has to be an integer.
The presence of matter fields however, fermions and vector bosons
with a Chern-Simons term, gives rise to an extra effective contribution
to the Chern-Simons charge $q_{gr}$.
\beq
q^{ren}_{gr} = q_{gr}
+\frac{1}{8} N_{g}~sign(m_{g}) -\frac{1}{16} N_f~sign(m_f)
\eeq
where $N_{g}$ is the number of vector bosons with 
topological mass $m_{g}$ and $N_f$ is the number of 
fermions of mass $m_f$. It is important that the corrections
are only dependent on the sign of the mass and not its absolute value.
This means that also at zero mass an effect is present. Within the purely
three dimensional case one speaks therefore of a parity anomaly, since
the basic tree level Lagrangian does not violate parity. Embedding the theory in
four dimensions with a preferred direction it is easy to understand that the sign is
important, since the sign of the mass in the Chern-Simons like term is fixed when
one chooses an orientation for the coordinate basis vectors.
We now assume that the fundamental  gravitational
laws have no preferred direction, implying $q_{gr}=0$.
The complete effective Chern-Simons term is then induced by the matter fields.
In this case the quantization condition gives rise to the following
identity
\begin{equation}
N_f \mp 2 N_g = 0~~{\rm mod} (16)
\end{equation}
whereby the minus sign is to be taken when the fermions and the bosons have the
same sign of the mass.
It is assumed that the fermions separately and the bosons separately have the same sign
for the mass, which is a reasonable assumption when they are part of
the same multiplets in a unified theory, since otherwise one would break the
gauge symmetry.
We see that the condition (7) is fulfilled for the vector bosons by themselves
 if the gauge group is $SU(5)$, giving $N_g=24$ and also for  the fermions
by themselves, when they
are in the $16$-dimensional spinor representation of $SO(10)$.
Moreover
it is desirable that the effective renormalized gravitational Chern-Simons
charge $q^{ren}_{gr}=0$, since otherwise it is difficult to understand
that the late universe is even approximately isotropic, because the gravitational
field equations themselves would have a preferred direction.
This condition is fulfilled if there are three generations of fermions
$3\times 16 - 2\times 24 = 0$.\\

Since ultimately the symmetry of $SU(5)$ gets broken one can wonder if the
consistency condition might play a role in the symmetry breaking pattern.
One would expect different signs for the subgroups $SU(3), SU(2), U(1)$
and the different representations of the fermions under the $SU(5)$ decomposition
$16= 10 + \bar 5 + 1$.

If we take
\beq
SU(3)\rightarrow +,\hskip 0.3cm SU(2)\rightarrow -,\hskip 0.3cm U(1)\rightarrow +
\eeq
and
\beq
10\rightarrow +,\hskip 0.6cm \bar 5\rightarrow -,\hskip 0.6cm 1\rightarrow -
\eeq
we get $2\times(8-3+1) - 3\times(10-5-1) = 0$.
Therefore the chain $SU(5)\rightarrow SU(3)\times SU(2) \times U(1)$
appears possible, however for the chain $SU(5)\rightarrow SU(4) \times U(1)$
there is no solution, with the above conditions.\\

We see therefore that the phenomenologically desirable gauge groups
 and representations
are  well described by the above conditions. We take this as an indication
that the actual topology of space-time is of the form $M_3\times U(1)$.
The universe has become very large due to inflation, so
the topology is not directly visible in the present epoch. However a remnant
of the topology could be the existence of a preferred direction in the
structure of the microwave background. An anisotropy at the largest scales
may actually have been seen in the WMAP data \cite{wmap}. A recent analysis of
 inflation \cite{CMBan1,CMBan2} in  Bianchi-I models indicates that such an
 asymmetry could be explained in such a cosmology.\\

\section{Discussion}
In the above we have presented an argument that determines,
 or at least constrains, the gauge group and the
 representations for the fundamental forces in nature.
Let us compare our argument with previous attempts in this
direction in the literature. An earlier attempt to derive 
$SU(5)$ was made in the context of $N=8$ supergravity \cite{ellis}.
In order to derive an acceptable model, a  number of dynamical
assumptions regarding composite states had to be made.
With the realization that $N=8$ supergravity is not a
finite theory and therefore not suitable as a fundamental theory
for all forces, attempts along this direction have largely stopped.
Another attempt was to use string theory in the form of the 
heterotic string, which implies a gauge group
 $E(8)\times E(8)$ \cite{rohm}. The reasoning used is somewhat
similar to our case. The group is selected by the absence
of an anomaly, in this case the conformal anomaly 
on the world sheet. Subsequently reducing the group to
something closer to the standard model appeared possible, but rather 
complicated. With the realization, that string theory allows for many vacua
with different gauge groups, the idea of a unique group has been abandoned
in this approach. These two attempts are similar in that they are very ambitious.
The assumption is that one determines the unique form of fundamental dynamics
from a given mathematical structure, which should subsequently contain
the observed forces of nature.\\

Our attempt is much more modest. We basically combine three
apparently unrelated physical inputs
in a somewhat surprising way. The inputs are the behaviour of anisotropic
universes, the pattern of the quantum  numbers of fermions and vector bosons,
and the parity anomaly. All of these have been known
for a long time. As an interesting result we found an indication
that the universe should be anisotropic at early times. One can wonder how solid the
basis of the inputs is. The behaviour of anisotropic, but homogeneous universes has been
studied exhaustively and cannot be seriously doubted. The unification pattern
of the forces in $SU(5)$ and the fermions in $SO(10)$ is quite apparent and convincing.
However this pattern can be proven wrong through the discovery of new particles, for
instance a Z' boson, that would enlarge the rank of the gauge group. If this were to happen
it is quite difficult to fulfill the constraints that we imposed and would most likely 
prove the argument to be wrong. 
The restriction (7), in particular when one requires strict equality,
 can become quite restrictive, when it
is combined with other expectations one might have for a unified
field theory. For instance, assuming that the fermions
should be in the fundamental representation of a chiral-anomaly
free gauge theory, we only found $E(8)$ with two generations
of fermions as a simple example. But here the quantization 
condition is not satisfied for the generations separately,
but only in the presence of both of them.
There is somewhat of a controversy\cite{percacci} regarding the
quantization condition in formula (5), since strictly speaking this condition uses
 a Euclidean formulation for gravity\cite{hawking}. This is however a general problem in quantum 
gravity. It is a difficult and unresolved problem within
quantum gravity whether a Euclidean formulation is necessary. In ordinary quantum field
theory one has of course the Euclidicity postulate, that the Minkowski space
Green's functions should be analytic continuations of Euclidean Green's functions,
in order to satisfy causality. 
A last worry is the question whether the gravitational parity
anomaly that exists in three dimensions can be lifted to four dimensions as argued in the
text. For the Yang-Mills case this was worked out in detail in\cite{frans1}, with the result
that there is a direct relation between the three dimensional
 parity anomaly and the four dimensional
CPT anomaly. This derivation can directly be applied to gravity,
 at least in its linearized form, since in this form
there is no fundamental difference between the spin connection and an ordinary gauge field.\\

Finally we consider the question, whether the inputs can be relaxed and whether 
extra conditions could still be present. 
The way the argument was presented, the actual structure of the early
universe plays an essential role. This condition might be
relaxed within the context of quantum gravity, where
topology change is presumably possible. Within a quantum
gravitational context it would therefore not
be necessary that the actual topology of space-time be $M_3\times U(1)$,
but it would be sufficient that this topology is potentially possible.
This might lead to  possibly stronger conditions, if one imposes, that
the spectrum of the matter fields should be such that the theory is anomaly
free for all possible compactifications of the actually existing space-time.
We wish to point out that our approach to symmetry has
some similarity to the one in \cite{nielsen}. \\

%\section{Bibliography}

\end{document}